*The Metropolis algorithm: A useful tool for epidemiologists*

**Authors:**
Alexander P Keil, Jessie K Edwards, Ashley I Naimi, Stephen R Cole

**Acknowledgments**
1) Author affiliations: Department of Epidemiology, University of North Carolina, Chapel Hill, North Carolina (Alexander P. Keil, Jessie K Edwards, Stephen R Cole); Department of Epidemiology, University of Pittsburgh, Pittsburgh, PA, USA (Ashley I Naimi)
2) Author contributions: Dr. Keil was responsible for conception and design, analysis and drafting of the manuscript. Dr. Edwards was responsible for conception and design and drafting of the manuscript. Dr. Naimi was responsible for conception and design, and drafting of the manuscript. Dr. Cole was responsible for conception and design, and drafting of the manuscript.
3) Funding: This work was supported by the National Institutes of Health (Grant R01 ES029531).
4) Conflict of interest statement: the authors have no conflicts of interest to declare.

ABSTRACT

The Metropolis algorithm is a Markov chain Monte Carlo (MCMC) algorithm used to simulate from parameter distributions of interest, such as generalized linear model parameters. The "Metropolis step" is a keystone concept that underlies classical and modern MCMC methods and facilitates simple analysis of complex statistical models. Beyond Bayesian analysis, MCMC is useful for generating uncertainty intervals, even under the common scenario in causal inference in which the target parameter is not directly estimated by a single, fitted statistical model. We demonstrate, with a worked example, pseudo-code, and R code, the basic mechanics of the Metropolis algorithm. We use the Metropolis algorithm to estimate the odds ratio and risk difference contrasting the risk of childhood leukemia among those exposed to high versus low level magnetic fields. This approach can be used for inference from Bayesian and frequentist paradigms and, in small samples, offers advantages over large-sample methods like the bootstrap.

The Metropolis algorithm is used to simulate values from distributions of interest (1). Such distributions can be simple, as in the case of sampling from a bivariate normal distribution. As recent history in Bayesian statistics has shown, such distributions can be more complex as in the case of estimating posterior distributions of parameters from a logistic regression model (2). The Metropolis algorithm is foundational within the broader family of Markov chain Monte Carlo (MCMC) methods. Key aspects of the Metropolis algorithm are present in modern, efficient MCMC methods such as Hamiltonian Monte Carlo, so understanding this algorithm provides a gateway to understanding MCMC (3).

The Metropolis family of MCMC methods has been popularized through Bayesian statistics (4) but is a tool that has broader utility. A central contrast with the more familiar maximization algorithms, like maximum likelihood estimators (which typically use some version of Newton's algorithm), is that the Metropolis algorithm simulates a set of values from a distribution, rather than returning a single value, as in maximum likelihood estimation (5).

Here, we provide a step-by-step description of the Metropolis algorithm in the familiar setting of estimating the odds ratio and risk difference from logistic regression. We demonstrate the use of the Metropolis algorithm in analyzing data from a study of magnetic fields and childhood leukemia. We show that the Metropolis algorithm can be used for frequentist or Bayesian inference and serve as a flexible tool for quantifying uncertainty in settings where bootstrap approaches may fail.

**TARGET PARAMETERS, TARGET DISTRIBUTIONS**

Target parameters (6) refer to a mapping of model parameters into a possibly smaller number of parameters that characterize the object of interest. In contrast, the "target distribution," following standard MCMC terminology refers to the joint distribution over all of the parameters (including latent variables) that are necessary in a model, including nuisance parameters that are not of direct scientific interest. For example, our target parameter might be the (causal) risk difference contrasting low versus high exposure to magnetic fields, computed via simple operations on predictions from a logistic regression model (7). However, in the MCMC framework, our target distribution would be the joint distribution over all logistic model parameters used to estimate the odds of childhood leukemia at a given level of exposure to magnetic fields (and values of confounders or other control variables, as applicable) (8).

**MONTE CARLO AND THE FIRST MCMC METHOD: THE METROPOLIS ALGORITHM**

The Metropolis algorithm belongs to a large class of methods that rely on simulation, known as Monte Carlo methods (9–11). Applications of Monte Carlo methods include uncertainty propagation in bias analysis (12,13), confidence interval estimation (14), multiple imputation (15), mathematical modeling of infectious diseases (16,17), and mathematical integration of complex functions (18). The latter application is particularly useful within modern causal inference methods to estimate average causal effects (19). Other uses exist (20).

A basic form of Monte Carlo estimation relies on simulating an <u>independent</u> sequence of values directly from the target distribution or from a distribution that is "close" to the target distribution, after which some samples can be rejected for failing to conform to the target distribution (21). These methods can be impractically inefficient in when the target distribution is a function of many parameters because the target distribution becomes increasingly difficult to simulate from directly or approximate well.

MCMC is one approach to making simulations more computationally efficient for purposes of estimating and quantifying uncertainty. MCMC uses the simulation-based approach of Monte Carlo, but it simulates a sequence of values where the simulation of each value takes into account the previous simulated value (22). Loosely, MCMC leverages serial dependence to "linger" in high probability regions of the target distribution when drawing new samples. Markov chain theory provides general conditions under which MCMC algorithms sample from a specific target distribution (22). Under Markov chain theory, MCMC can be used to sample from a variety of distributions, given only partial knowledge of those distributions. Specific MCMC algorithms differ in what types of partial knowledge are necessary.

The first MCMC algorithm, known as the Metropolis algorithm, was developed in the 1950s to study states of physical matter (1). In spite of its age, the Metropolis algorithm continues to have many uses because it is general, yet relatively simple. Major elements of the algorithm are also present in modern MCMC methods (9). The

Metropolis algorithm is thus a useful tool in its own right, and understanding it is a prerequisite to a deep understanding of modern MCMC.

Using the Metropolis algorithm to sample from a distribution requires only that we are able to evaluate, up to a proportionality constant, the probability density of the target distribution at specific parameter values (i.e., the density at a parameter value relative to the density at other values). We show, for example, that the Metropolis algorithm can be used for logistic regression, provided only that we can calculate a binomial likelihood. Further, we show that one can also easily obtain point and uncertainty estimates for other target parameters, given this model, such as risk differences.

**A MOTIVATING EXAMPLE: MAGNETIC FIELDS AND CHILDHOOD LEUKEMIA**

Our study data were collected as part of a case-control study of childhood leukemia and magnetic field measurements (8) and consist of 36 leukemia cases and 198 controls with magnetic field measurements and are provided in **Table 1**. We assume a logistic model given by **Equation 1.**

$$P(Y = 1|X, \beta_0, \beta_1) = \frac{\exp(\beta_0 + \beta_1 X)}{1 + \exp(\beta_0 + \beta_1 X)} = \text{expit}(\beta_0 + \beta_1 X), \qquad [1]$$

where $Y = 1$ indicates a case of childhood leukemia, $X = 1$ indicates exposure to high magnetic fields ($> 0.3 \ \mu T$) versus low magnetic fields ($\leq 0.1 \ \mu T$), and $\beta_0, \beta_1$ are the parameters of interest. Because we do not consider potential confounders in this model, the joint distribution of $\beta_0, \beta_1$ is the target distribution for this problem. For simplicity, we will denote the collection of $\beta_0, \beta_1$ as $\beta$, where appropriate. In a case control study, $\beta_0$ is partly determined by the ratio of cases to controls and is thus a design feature, but we leverage this parameter below to estimate more meaningful quantities.

Given this model and the data from **Table 1**, we can sample from a joint posterior probability distribution of $\beta$, using MCMC. By "sampling," we mean simulating, or "drawing" values from the distribution. Given samples from this joint probability distribution, which comprise a numerical approximation of the target distribution, we can easily estimate the marginal distributions of $\beta_0$ and $\beta_1$, which can be used for point estimation (e.g. taking the mean of the samples) and uncertainty quantification (e.g. taking standard deviations or quantiles of the samples). Thus, we can think of an the samples as samples from the posterior distribution that integrates our imperfect prior knowledge and data regarding $\beta$ (as a Bayesian might), or merely consider these samples to be a source for summary point estimates for $\beta$ (as a frequentist might).

**THE METROPOLIS ALGORITHM: RECIPE**

As we noted, MCMC comprises a family of methods to draw serially-correlated samples from a target distribution, which we denote as $h$. For our example, $h(\beta|Y,X)$ is the joint density function of the parameters of a logistic model for the log-odds of childhood leukemia (**Equation 1**), given the data. Using the Metropolis algorithm to sample $\beta$ requires two pieces. First, we need a function that calculates a value proportional to the target distribution at a given value of $\beta$. We denote this function by $m(\beta|x,y) = L(\beta|x,y)\pi(\beta) \propto h(\beta|x,y)$, where $L(\beta|x,y)$ and $\pi(\beta)$ are the likelihood and prior probability density functions, respectively. This is similar to the usual expression of

Bayes theorem ($P(B|A) = P(A|B) * P(B)/P(A)$), but the denominator (a marginal density function of the data) is omitted because it cancels out in all calculations of the Metropolis algorithm. We use a uniform prior, which allows us to simplify the function to equal the likelihood (explained further below). Second, we need a "proposal distribution" which is used to create candidate values or "guesses" at new draws of $\beta$. These candidate values are sequentially accepted or rejected to create a "chain", which is a series of values of $\beta$, (i.e., the sequences $\beta_0^{(1)}, \beta_0^{(2)}, \ldots, \beta_0^{(M)}$ and $\beta_1^{(1)}, \beta_1^{(2)}, \ldots, \beta_1^{(M)}$, where the superscript (1), (2), … (M) denotes the iterations of the Metropolis algorithm, as explained further below).

We must first initialize each sequence of parameter values. For our example these initial guesses are intentionally poor to emphasize convergence to the target distribution. We used $(\beta_0^{(1)}, \beta_1^{(1)}) = (2, -3)$, which imply a baseline odds of leukemia of 7.3 in the study sample and sample odds ratio of 0.05. These are not prior values and their choice, as long as they fall within the space covered by the algorithm, should only affect the number of iterations until the algorithm has converged to the target distribution.

Then, in turn, the two core steps described below are alternated until we reach a predefined number of samples ($M$, using our notation above). The number of samples is chosen to be large, say 5,000 or 100,000, with attention to the fact that these samples are not independent because the proposed values of $(\beta_0^{(j+1)}, \beta_1^{(j+1)})$ are always centered at $(\beta_0^{(j)}, \beta_1^{(j)})$, as we will see below. We discuss implications for this dependence below when detailing how to choose $M$. We present pseudo-code (i.e. human readable code that doesn't correspond to a particular programming language) for the algorithm in **Figure 1**, with a deeper explanation in the following steps.

**(FIGURE 1 HERE)**

**Step 1: Generate candidate values of parameters from the proposal distribution**

A proposal distribution at iteration $j + 1$ yields *candidate* values for $\beta^{(j+1)}$, which are denoted by $c$. For our example, we are estimating $\beta_0$ and $\beta_1$, therefore $c$ will be a collection of two candidate values, $c_0$ and $c_1$. In the Metropolis algorithm, the proposal distribution, denoted by $q$, must be symmetric (i.e. $q(a|b) = q(b|a)$ for all values $a$ and $b$). We use a normal proposal distribution, which is a commonly used symmetric distribution. For our example at iteration 2, the candidate $c$ is generated by simulating a value from $q(c|\beta^{(1)}) \sim N_2(\beta^{(1)}, \Sigma_c)$, which denotes a bivariate normal distribution with mean vector $(\beta_0^{(1)}, \beta_1^{(1)})$ and covariance matrix $\Sigma_c$. While we discuss below how $\Sigma_c$ can be selected to enhance efficiency, for our example we use an arbitrarily selected covariance matrix, given by

$$\Sigma_c = \begin{bmatrix} 0.1 & 0.0 \\ 0.0 & 0.1 \end{bmatrix}$$

For $j + 1 = 2$ we drew values of $c_0$ = 1.64, and $c_1$ = -2.78 as candidates for $\beta_0^{(2)}$ and

$\beta_1^{(2)}$, respectively.

**Step 2: The "Metropolis step" to accept or reject jumps to the candidate values**

The candidate values are either accepted or rejected (together as a set, in this example) in proportion to the relative values of $h(\beta|x,y)$ evaluated at $\beta = c$ vs. $\beta = \beta^{(j)}$. The "acceptance probability" for candidates at iteration $j + 1$, $pA^{(j+1)}$, is given in **Equation 2.** Recall that $h(\beta|x,y)$ need only be proportional to the likelihood/posterior of interest because proportionality constants cancel out in the acceptance probability calculation.

$$pA^{(j+1)} = minimum\left(1, \frac{h(c|x,y)}{h(\beta^{(j)}|x,y)}\right) \qquad [2]$$

In our example, the denominator is the product of independent binomial likelihoods under a logit link given by $h(\beta^{(j)}|x,y) = \prod_i^N \mu_i^{y_i}(1-\mu_i)^{y_i}$, where $\mu_i = \text{expit}\left(\beta_0^{(j)} + \beta_i^{(j)} x_i\right)$, and "expit" is defined in **Equation 1**. Note that there is no explicit prior $\pi(\beta^{(j)})$ in this expression, but the implied prior is uniform such that all values of $\beta$ are equally probable, so that $\frac{\pi(c)}{\pi(\beta^{(j)})} = 1$ and can be omitted. When proposed values increase the posterior probability relative to the previous iteration, they are always accepted. The use of the minimum function in **Equation 2** ensures that the acceptance probability does not exceed 1. We use log transformations when calculating the acceptance probability to avoid numerical issues. This quantity is readily calculable in statistical software.

In our example at $j = 1$, given our values of $\left(\beta_0^{(1)}, \beta_1^{(1)}\right) = (2, -3)$, and $(c_0, c_1) =$

$(1.64, -2.78)$, the log acceptance probability evaluates to $min(0, -361 - (-420)) = min(0, 59) = 0$, so the probability of accepting the candidate values is $\exp(0) = 1$. Thus, we accept the proposal and set $\left(\beta_0^{(2)}, \beta_1^{(2)}\right) = (1.64, -2.78)$, and the chains for $\beta_0$ and $\beta_1$ are now (2.00, 1.64) and (-3.00, -2.78), respectively. If we had rejected the candidate value, then we would set $\beta^{(j+1)} = \beta^{(j)}$.

The acceptance or rejection of four proposed jumps is demonstrated for iterations 253-256 of our algorithm (**Figure 2**), where high acceptance probabilities (Panels A and C) are both accepted updates, and low probability moves are both rejected (Panels B and D).

**(FIGURE 2 HERE)**

After iterating core steps 1 and 2 above $M$ times, we can make inferences. For a large Markov chain resulting from the Metropolis algorithm detailed above, inference proceeds as though all values of $\beta$ are simulated values from the target distribution. For example, the sample mean and standard deviation of the $M$ samples estimates the posterior mean and posterior standard deviation. Credible intervals can be constructed by taking percentiles of the $M$ samples (or via many other methods (23)).

**SIMULATION PRECISION: HOW MANY SAMPLES ARE NEEDED?**

The number of simulated samples, $M$, is can be considered in an analogy with sample size in an experiment: as sample size grows, the precision around our estimates also grows. Because we can choose $M$, this means that the simulation estimate of the target distribution can be made arbitrarily precise by making $M$ extremely large. Sampling precision refers to the how much our simulated distribution changes as we draw additional samples. That precision gives us a sense of whether or not we are within an acceptable margin of simulation error for reporting estimates. Unlike a sample of independent observations, our Metropolis estimates are positively correlated along the chain, reducing the information provided by additional samples.

One intuitive way to estimate sampling precision is through the "effective sample size" of the mean of a Markov chain, given by $M'(\beta) = \frac{M}{1+2\sum_{\ell=1}^{\infty} \rho_\ell(\beta)}$, where $\rho_\ell(\beta)$ is the correlation in the values of $\beta$ between draws at iterations $j$ and iterations $j - \ell$ for all $j > \ell$ (also known as the lag-$\ell$ autocorrelation.) Notably, if all lagged autocorrelations are zero (as expected in a simple random sample from the posterior), then the effective sample size would equal the number of iterations. Thus, effective sample size quantifies how precise the estimate of the posterior mean is, expressed as a number of independent random samples (24,25). In practice, effective sample size must be estimated via a small sum of estimated autocorrelations, and effective sample size will generally be lower for quantities other than the posterior mean (see (26) for details).

Effective sample size can be used to estimate Monte Carlo standard error (denoted by $\hat{\sigma}_{E(\beta)}$), which can then be used to select the appropriate number of iterations. One standard approach to estimating the Monte Carlo standard error of the mean is to take the estimated standard deviation of $\beta$ and divide by the square root of the estimated effective sample size at the mean. Monte Carlo standard error can be used to estimate, for example, the probability that the estimated posterior mean is within a certain margin of error of the true posterior mean.

Effective sample size is intuitive to those familiar with simulation-based approaches, but other diagnostics can be used to assess whether moments of the distribution have "converged," meaning their estimates have a high probability of being within some small neighborhood around the true distribution moments. In simple settings, such as ours, convergence can be assured by just obtaining many effective samples, but in more complex settings, statistical diagnostics are available and should be used (e.g. chapter 11 of (4)).

**IMPROVING THE METROPOLIS ALGORITHM**

Above, we described the "random walk" Metropolis algorithm which is so named because the proposal distribution allows jumps of any length in any direction. First, this may not be ideal because random walks are equally likely to attempt to move the chain to high probability regions as they are to move to low probability regions (cf, e.g., Figure 2, Panel A versus Panel D). Second, the sampling efficiency is sensitive to size of the jumps of the (investigator specified) proposal distribution: jumps that are too small or too large result in high autocorrelation along the chain, which reduces effective sample size. As an analogy, we can think of a posterior distribution roughly as a "mountain" and the Metropolis algorithm as a procedure to map the terrain of the mountain. To explore a mountain efficiently, we need to ensure that we visit all sides of the mountain and avoid spending too much time in any particular spot. The random walk-procedure is analogous to an inefficient exploration technique of wearing a blindfold while alternately crawling or taking huge (super-hero sized) jumps.

Following Gustafson (27), we highlight solutions to both of these exploration problems: guiding and adapting. **Guiding** means we force all jumps of a component of $\beta^{(j)}$ to $\beta^{(j+1)}$ be in the same direction until a candidate is rejected. For example, for $\beta_1^{(j)}$, the proposal $c_1$ is constrained to be positive until a candidate is rejected (the initial sign of each proposal is randomly chosen at the first iteration), at which point $c_1$ is constrained to be negative until the next rejection, and so on (27,28). Guiding speeds up initial convergence and encourages full exploration of the target distribution. Analogously, a mountain would be fully explored more quickly by removing our blindfold and prioritize climbing up the mountain to see the other side. **Adapting** means that, during a burn-in period, we frequently update the proposal distribution covariance matrix $\Sigma$ to be proportional to an estimate of the covariance matrix of the target distribution (29,30). This changes the jump sizes. Analogously, adapting ensures that in our mountain exploration we do not take such large jumps that we jump into a neighboring valley, nor do we crawl up its side, both of which would slow exploration. Rather, we find, through feedback from the terrain, a good gait that allows us to explore the mountain effectively. Guiding and adapting can be combined.

EXTENDING BEYOND THE FIRST STEPS

We demonstrate performance of three Metropolis algorithms (random walk, guided, and guided and adaptive) using our study data and the model in **Equation 1.** We use maximum likelihood and the Metropolis algorithm to estimate a log-odds ratio ($\beta_1$) contrasting exposure to between cases and controls. Additionally, we estimate population risk differences from these case-control data. Estimation of risk differences is not possible from case-control data without additional information. Therefore, we use a population leukemia incidence of 4.8/100,000 child-years (31), which together with the (appropriately sampled) case-control data allows computation of risk differences (32,33). As an aside, we note that this approach is a simple form of Bayesian g-computation (34). The Metropolis algorithm was used with uniform $(-\infty, \infty)$ priors and the null-centered normal priors used by Cole et al with these data (21)). While uniform priors are common (and used didactically here), such priors are not uninformative (35) and lack benefits of well-chosen, informative prior distributions.

As shown in **Table 2**, MCMC and maximum likelihood estimates were similar under uniform priors, especially using the posterior mode (estimated by explicitly evaluating the posterior distribution over the MCMC samples and known as the "maximum a posteriori" estimate). The risk difference using the median of the simulated distribution was similar to the estimate from maximum likelihood. Notably, bootstrapping failed to produce confidence intervals for the maximum likelihood method (due to quasi-complete separation in 8/200 resamples) demonstrating an advantage of MCMC in small data sets. Informative priors reduced posterior variance, and yielded estimates for $\beta_0, \beta_1$ similar to those reported by Cole et al (21). These priors were heavily informative based on the difference between the two sets of MCMC estimates. The guided and adaptive algorithm was the most computationally efficient algorithm, based on effective sample size (**Table 2**). This result agrees with visual evidence from trace plots, where the guided and adaptive algorithm moved quickly towards the posterior means and demonstrated frequent traversals of the posterior, implying low autocorrelation and good sampling efficiency (**Figure 3**). Replication code and movies that demonstrate the movement of each algorithm through the posterior are available at <author-Github-page>.

DISCUSSION

Modern MCMC methods, like those used by STAN, R, SAS, and other software programs, use evolved aspects of the Metropolis algorithm. Metropolis steps have long been integral to extensions of Gibbs samplers (36) when a given parameter may not be amenable to efficient Gibbs sampling (giving rise to "Metropolis-within-Gibbs" samplers (22)). Further, basic extensions of the "Metropolis step" appear in efficient, modern MCMC implementations such as Metropolis-Hastings (which allows asymmetric proposals, e.g. (27)) Hamiltonian Monte Carlo (3), reversible jump Monte Carlo (37) and Newtonian Monte Carlo (38). Adaptative proposals and random-walk suppression (like guiding) also show up in modern MCMC. Compared with modern MCMC, the original Metropolis algorithm is computationally simpler, general, and reasonably efficient for many data analytic problems. The Metropolis algorithm thus remains useful as analytic tool in its own right and also as a learning tool for understanding more advanced MCMC approaches.

The ubiquity of the Metropolis step in MCMC raises the basic question: *why does the Metropolis algorithm work at all*? A technical answer to this question with references is given by Gelman et al (section 11.2 of (4)). Intuition can be gained by exploring the Metropolis algorithm in a simple setting of a posterior of one parameter $\theta$ with two possible values $a$ and $b$ and a proposal distribution that always proposes to switch between these two values (implying symmetry). Suppose that we know that the probability mass distribution of $\theta$ is given as $Pr(\theta = b) = 2 \times Pr(\theta = a)$, or, in words, the probability mass at $b$ is twice what it is at $a$. This 2:1 ratio is the basic quantity needed in the Metropolis algorithm. Since the total probability of $\theta = a$ or $\theta = b$ must sum to 1, we can directly calculate $Pr(\theta = b) = \frac{2}{3}, Pr(\theta = a) = \frac{1}{3}$, so we must show that the Metropolis algorithm converges to these values.

The probability of the chain moving from $b$ to a value with a lower posterior probability, $a$, is equal to $Pr(\theta = a)/Pr(\theta = b)$. Thus, the acceptance probability for moving from $b$ to $a$ is (1/3)/(2/3) = 0.5. Say a Markov chain is initialized so that the first iteration value is set to $\theta^{(j)} = b$. For subsequent draws, we will always jump from $\theta^{(j)} = a$ to $\theta^{(j+1)} = b$, and we will jump from $\theta^{(j)} = b$ to $\theta^{(j+1)} = a$ with probability 0.5. Thus, there is only one possible way to draw $\theta^{(j+1)} = a$, where $Pr(\theta^{(j+1)} = a) = 0.5 \times Pr(\theta^{(j)} = b)$; However, there are two possible ways to draw $(\theta^{(j+1)} = b)$, where $p(\theta^{(j+1)} = b) = 1.0 * Pr(\theta^{(j)} = b) + 0.5 * Pr(\theta^{(j)} = b)$. Because the probabilities come from the same distribution, regardless of the sequence in the chain, we also know that $p(\theta^{(j)} =$

$b$)= $Pr(\theta^{(j+1)} = b) = Pr(\theta = b)$. That is, the probability of drawing $B$ at any given iteration is the same across all iterations, irrespective of the prior states of the chain. Substituting gives us $0.5 * Pr(\theta^{(j)} = b) + Pr(\theta^{(j+1)} = b) = 1.5 * Pr(\theta^{(j)} = b) = 1.0$, and $Pr(\theta^{(j)} = b)$=2/3, meaning that $Pr(\theta^{(j)} = b)$=1/3. Thus, the samples drawn at any given iteration represent draws from the target distribution $h(\theta)$.

The acceptance probability is 1 every time a proposed value of the chain $c$ has a higher posterior density than the current value $\beta^{(j)}$. Thus, even for starting values of a chain that are far from the center of the posterior distribution, sampled values always tend towards the high probability regions of the posterior. Thus, the algorithm will concentrate around the mode of the distribution, even if starting values are very far from this mode and will (loosely speaking) draw samples that are representative of the target distribution.

The same basic calculations from a parameter with two possible values can be easily extended to one with three or four possible values and so on, and the intuition straightforwardly generalizes to continuously valued parameters, as in our worked example.

While MCMC has become synonymous with Bayesian inference, it can be also be used for frequentist inference. Using Bayesian algorithms for frequentist inference is sometimes called "frequentist pursuit" because it pursues frequentist goals such as unbiased point estimation and unbiased confidence interval coverage (39). The Metropolis algorithm yields a parameter distribution, given a fixed set of data. Conversely, frequentist variance estimates derive from the view that the data are not fixed but are sampled from some larger population. Thus, the Metropolis algorithm does not explicitly address this sampling variation, unlike frequentist estimates. However, our analysis shows that estimates can be similar in small samples (40), and inference converges in large samples for many classes of statistical models (but not all (41)).

MCMC algorithms are powerful, but in practice currently fill only a small, Bayesian niche in the epidemiologic literature. Our example demonstrates one scenario where default frequentist approaches using the bootstrap can fail to provide useful estimates of uncertainty, but MCMC methods succeed. We provide freely available code, demonstrating the relative ease of implementing custom MCMC routines. Our hand-coded approach provides clarity for the algorithms underlying the methods and, we hope, opens up the black box of MCMC for analysts that may find it useful for inference.

# TABLES

**Table 1. Data for case-control study of high versus low magnetic field measurements among 36 leukemia cases and 198 controls**

|  | Magnetic field | |
|---|---|---|
|  | ≤ 0.1 μT | > 0.3 μT |
| Controls | 193 | 5 |
| Cases | 33 | 3 |

**Table 2. Performance of maximum likelihood versus MCMC: Odds ratio and risk difference estimates for high versus low magnetic field measurements among 36 leukemia cases and 198 controls**

| | $\hat{\beta}_1$[a] | | | | | | | | |
|---|---|---|---|---|---|---|---|---|---|
| **Method** | Mean | Median | Mode | SD[b] | OR[c] | 95% CI/PI[d] | RD[e] | 95% CI/PI[d] | ESS[f] |
| **Maximum Likelihood** | | | 1.26 | 0.75 | 3.51 | (0.69, 15.4) | 0.11 | [i] | |
| **Metropolis-MCMC, Uniform prior[g]** | | | | | | | | | |
| Random Walk | 1.23 | 1.25 | 1.26 | 0.79 | 3.49 | (0.67, 15.0) | 0.11 | (-0.02, 0.59) | 1929 |
| Guided | 1.19 | 1.22 | 1.25 | 0.80 | 3.38 | (0.64, 15.0) | 0.11 | (-0.02, 0.60) | 15309 |
| Guided, Adaptive | 1.20 | 1.22 | 1.26 | 0.80 | 3.40 | (0.63, 15.2) | 0.11 | (-0.02, 0.60) | 34680 |
| **Metropolis-MCMC, Normal Prior[g,h]** | | | | | | | | | |
| Guided, Adaptive | 0.53 | 0.54 | 0.55 | 0.55 | 1.71 | (0.57, 4.91) | 0.03 | (-0.02, 0.18) | 40769 |

[a] Log-odds ratio for association between X and Y. Given as maximum likelihood estimate (under mode) or the sampling distribution mean, median, or mode from MCMC
[b] Standard error of maximum likelihood estimate or standard deviation of sampling distribution from MCMC
[c] Odds Ratio, $\exp(\hat{\beta}_1)$ using maximum likelihood estimate or median of sampling distribution from MCMC
[d] 95% Confidence interval based on normal approximation for maximum likelihood or 2.5th and 97.5th percentiles of sampling distribution from MCMC
[e] Risk difference*1,000 estimate using model offset (7) with logistic model (maximum likelihood) or median of sampling distribution from MCMC
[f] Effective sample size: estimate of sampling distribution information contained in Markov chain, expressed in equivalent sample size
[g] 100,000 iterations after 1,000 iteration burn-in
[h] Normal priors with means of 0 for $\beta_0$ and $\beta_1$ and variances of 100 ($\beta_0$) and 0.5 ($\beta_1$)
[i] Bootstrap confidence interval failed for risk difference due to logistic model separation in 8/1000 bootstrap iterations

## FIGURES

Figure 1. Pseudo-code describing the Metropolis algorithm to simulate from the likelihood of the model parameters exploring the association between childhood leukemia (Y) and exposure to high magnetic fields (X) in a case-control study, where the likelihood is given as proportional to the Bernoulli probability mass function (Bernoulli.PMF) and acceptance/rejection is based on a random draw from a Bernoulli distribution (Bernoulli.random).

---

For j = 1, …. M

1. Draw candidate values from proposal distribution:

   Draw $p(c|\beta^{(1)}) \sim N_2(\beta^{(1)}, \Sigma_c)$,

2. Calculate (log) acceptance probability $(pA^{(j+1)})$:

$$\ln(pA^{(j+1)}) = \text{MIN}\left(0, \sum_{i=1}^{N}\left(\ln(f(c|x_i, y_i)) - \ln\left(f(\beta^{(j)}|x_i, y_i)\right)\right)\right)$$

   Where
$$\ln(f(c|x_i, y_i)) = \ln\left(\text{Bernoulli.PMF}(y_i|mean = expit(c_0 + c_i\ x_i))\right)$$
   and
$$\ln\left(f(\beta^{(j)}|x_i, y_i)\right) = \ln\left(\text{Bernoulli.PMF}\left(y_i|\ mean = expit\left(\beta_0^{(j)} + \beta_1^{(j)} x_i\right)\right)\right)$$

3. Accept/reject candidate values:

   IF Bernoulli.random$(mean = pA^{(j+1)}) = 1$ : set $\beta^{(j+1)} = c$
   ELSE: set $\beta^{(j+1)} = \beta^{(j)}$

END

---

Figure 2. Details (iterations 253 to 256 shown in panels A to D) of the random walk metropolis algorithm for fitting a logistic model to the childhood leukemia study data. The contours represent the likelihood function (with the center as the maximum), round dots represent sampled points (with darker points having been sampled more than once), while the square points represent proposed points and the arrows the proposed move. The number within the box represents the acceptance probability $(pA^{(j+1)})$ for the proposed move, where the accepted jumps are represented in boxes **A** $(pA^{(254)} = exp(-100.25 - -100.18) = 0.94)$ **and C** $(pA^{(256)} = exp(-100.72 - -100.25) = 0.62)$ **and the rejected jumps in B** $(pA^{(255)} = exp((-101.51 - -100.25) = 0.28)$ **and D** $(pA^{(257)} =$

$exp(-103.07 - -100.72) = 0.10)$.

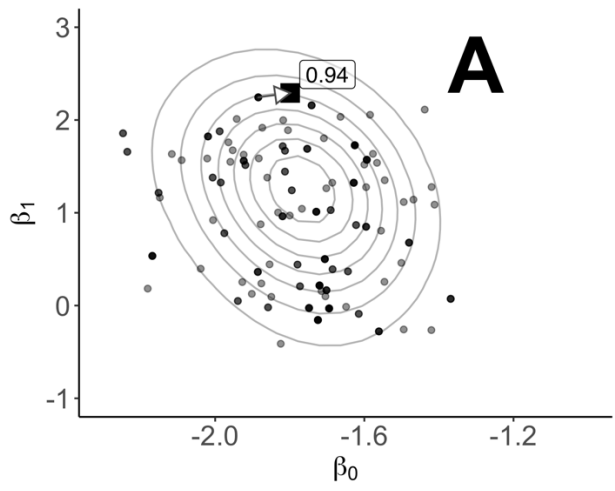
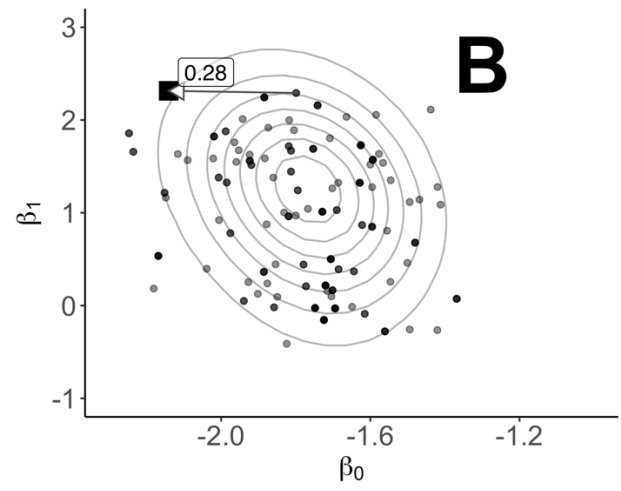
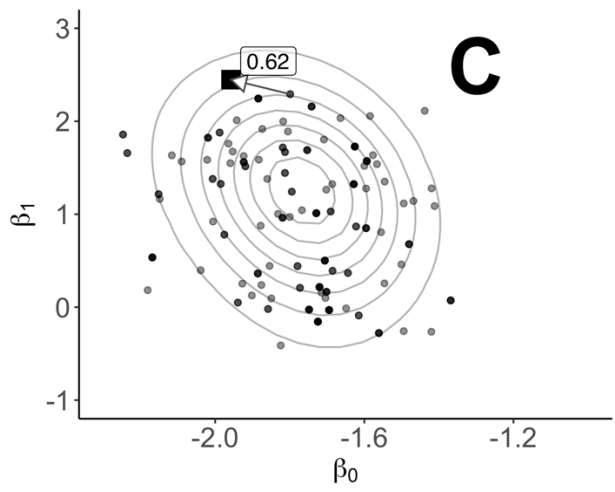
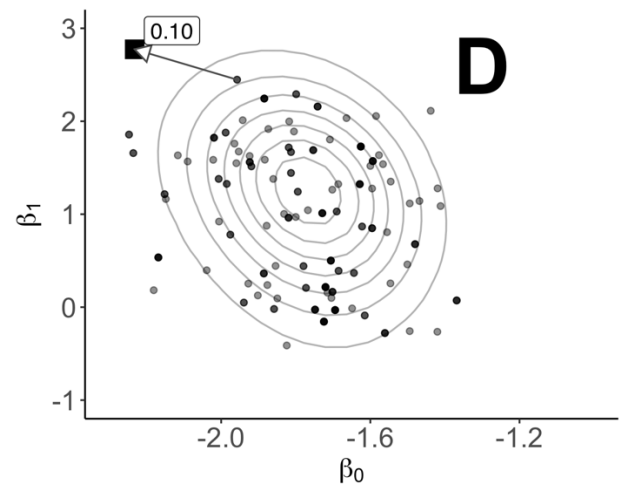

**Figure 3. Truncated trace plots demonstrating features of three variations of the Metropolis-Hastings algorithm for estimating $\beta_1$. The trace plots demonstrate that all variations sample the same region (panel A), and the plots demonstrate slow convergence (panel B) and slow traversal of the posterior (panel C) of the random walk Metropolis Hastings algorithm, with improved convergence for the guided, and guided, adaptive algorithms and faster traversal (rapid change of slope of change across iterations) for the guided, adaptive algorithm.**

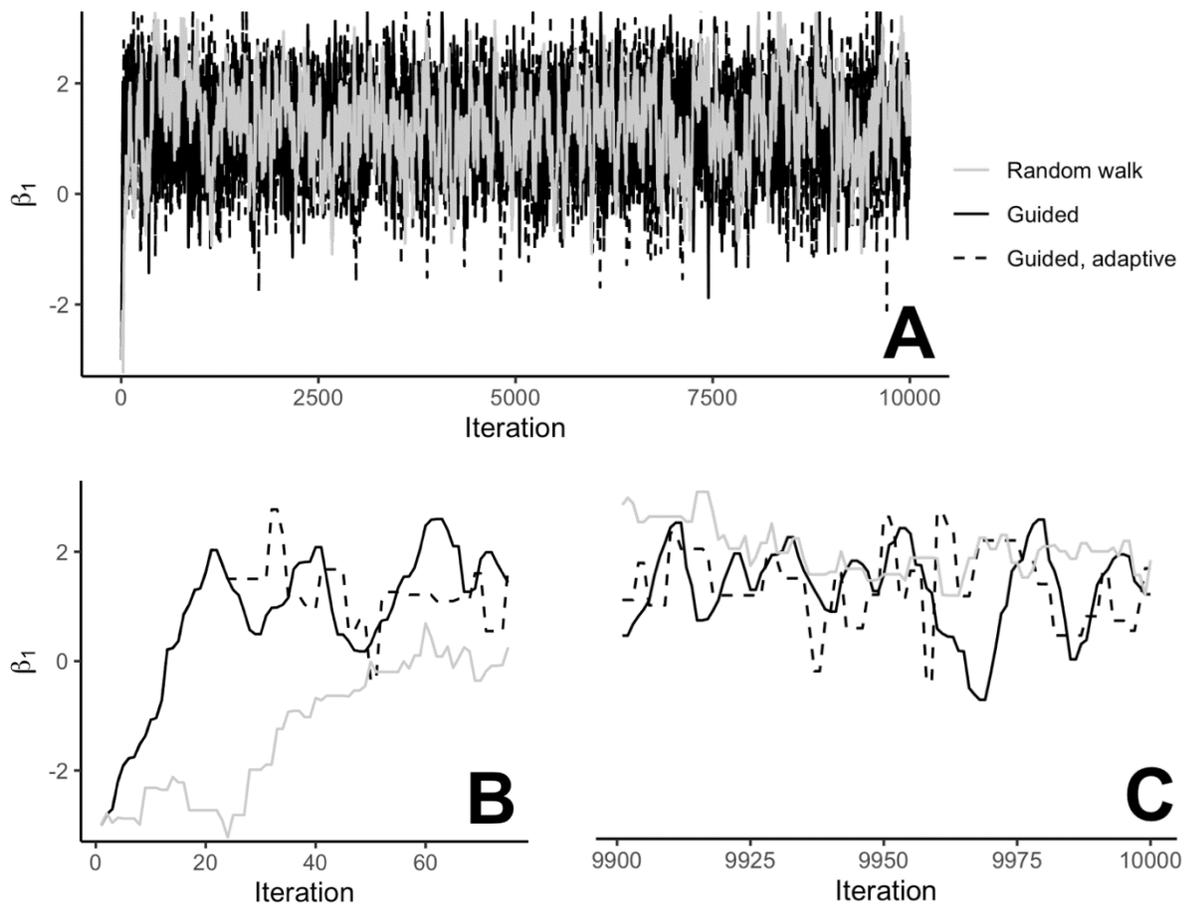